\documentclass[10pt,prl,aps,superscriptaddress,twocolumn,showpacs]{revtex4}
\usepackage{graphicx}

\usepackage{amssymb}

\begin{document}

\title{Emergence of U(1) symmetry in the 3D XY model with Zq anisotropy}

\author{Jie Lou}
\affiliation{Department of Physics, Boston University, 
590 Commonwealth Avenue, Boston, Massachusetts 02215}

\author{Anders W. Sandvik}
\affiliation{Department of Physics, Boston University, 
590 Commonwealth Avenue, Boston, Massachusetts 02215}

\author{Leon Balents}
\affiliation{Department of Physics, University of California, Santa 
Barbara, CA 93106-4030}

\begin{abstract}
We study the three-dimensional XY model with a $Z_q$ anisotropic term. At temperatures $T < T_{\rm c}$
this dangerously irrelevant perturbation is relevant only above a length scale $\Lambda$, which diverges as a power of 
the correlation length; $\Lambda \sim \xi^{a_{q}}$. Below $\Lambda$ the order parameter is U(1) symmetric. We derive 
the full scaling function controlling the emergence of U(1) symmetry and use Monte Carlo results to extract the exponent 
$a_q$ for $q=4,\ldots,8$. We find that $a_q \approx a_4 (q/4)^2$, with $a_4$ only marginally larger than $1$. We discuss 
these results in the context of U(1) symmetry at ``deconfined" quantum critical points separating antiferromagnetic and 
valence-bond-solid states in quantum spin systems.
\end{abstract}

\date{\today}


\maketitle

A salient feature of the recently proposed theory of "deconfined" quantum critical points, which separate N\'eel 
and valence-bond-solid (VBS) ground states of antiferromagnets on the square lattice, is the emergence of U(1) symmetry
\cite{senthil}. The VBS is either dimerized on columns or forms a square pattern with plaquettes of four strongly 
entangled spins \cite{levin,sachdevrmp}. In both cases there are four degenerate patterns and, thus, $Z_4$ symmetry is broken. 
However, as the critical point is approached the theory predicts a length scale $\Lambda$, diverging faster than 
the correlation length, $\Lambda \sim \xi^a$, $a>1$, below which the distinction between columnar and plaquette 
VBS states disappears. The nature of the VBS state is manifested only when coarse-graining on length-scales 
$l > \Lambda$, whereas for $l < \Lambda$ the $Z_4$ symmetry is unbroken and is replaced by an emergent U(1) 
symmetry characterizing the fluctuations between columnar and plaquette order.

Quantum Monte Carlo simulations \cite{sandvikvbs} of an $S=1/2$ Heisenberg model with four-spin couplings have
recently provided concrete evidence for a continuous N\'eel--VBS transition, and also detected U(1) symmetry 
in the VBS order-parameter distribution $P(D_x,D_y)$, where $D_x$ and $D_y$ are VBS order parameters for horizontal 
and vertical dimers. There is no trace of the expected $Z_4$ anisotropy in the VBS phase---the distribution 
is ring shaped---although the finite-size scaling of the squared order parameter shows that the system 
is long-range ordered. This can be interpreted as the largest studied lattice size $L=32 < \Lambda$. 
A ring-shaped distribution was also found in simulations of an SU(N) generalization of the $S=1/2$ Heisenberg 
model \cite{kawashima}---possibly a consequence of proximity of this system to a deconfined quantum-critical point. 

In order to better understand the U(1) features of these VBS states, and to guide future studies of them, we here
exploit a classical analogy. In the three-dimensional XY model including a $Z_q$-anisotropic term,
\begin{equation} 
{\cal H}=-J\sum_{(i,j)}\cos(\theta_i-\theta_j)-h\sum_{i}\cos(q\theta_i),
\label{hamiltonian}
\end{equation}
the anisotropy is dangerously irrelevant for $q\ge 4$ \cite{jose,blankschtein,caselle,oshikawa,carmona}, i.e., 
the universality class is that of the isotropic XY model but the perturbation is relevant for $T < T_{\rm c}$
above a length-scale $\Lambda$. In the closely related $q$-state clock model, the anisotropy is dangerously
irrelevant for $q\ge 5$. While numerical studies \cite{scholten,miyashita,hove} have confirmed the irrelevance 
of the anisotropy at $T_{\rm c}$, the associated $\Lambda$ has, to our knowledge, not been extracted numerically, 
except for an analysis of the 3-state antiferromagnetic Potts model, which corresponds to $Z_6$ 
\cite{oshikawa,aharony}. 

Here we report results of Monte Carlo simulations for $4\le q \le 8$ on periodic-boundary lattices with $N=L^3$ sites 
and $L$ up to $32$. In addition to  Metroplis single-spin updates, we also use Wolff cluster updates \cite{wolff} to reduce 
critical slowing down. We sample the order-parameter distribution $P(m_x,m_y)$, where
\begin{equation}
m_x=\frac{1}{N}\sum_{i=1}^N \cos(\theta_i),~~~m_y=\frac{1}{N}\sum_{i=1}^N \sin(\theta_i).
\end{equation}
The standard order parameter can be defined as
\begin{eqnarray}
\langle m\rangle & = & \int_{-1}^1 dm_x\int_{-1}^1 dm_y P(m_x,m_y)\left ( m_x^2+m_y^2 \right )^{1/2} \nonumber \\
& = & \int_0^1 dr \int_0^{2\pi} d\theta r^2P(r,\theta).
\label{mag}
\end{eqnarray}
We will compare this with an order parameter $\langle m_q\rangle$ which is sensitive to the angular distribution;
\begin{equation} 
\langle m_q\rangle = \int_0^1 dr \int_0^{2\pi} d\theta r^2P(r,\theta)\cos(q\theta).
\label{magstar}
\end{equation}

\begin{figure}
\centerline{\includegraphics[width=6.75cm,clip]{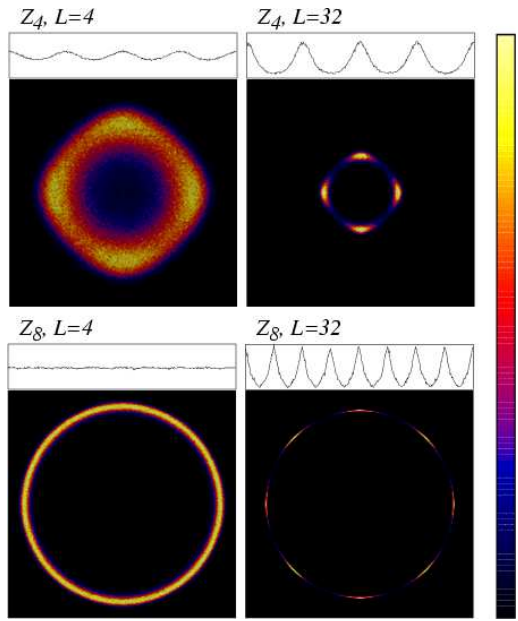}}
\caption{(Color online) $P(m_x,m_y)$ at $h/J=1$ for $q=4,8$, $L=4,32$.
The temperature $T/J=2.17$ for $Z_4$ and $1.15$ for $Z_8$; both less than $T_{\rm c}/J \approx 2.20$. The size of 
the histograms corresponds to $m_{x,y} \in [-1,1]$. Angular distributions $P(\theta)$ with $\theta \in [0,2\pi]$ 
are shown above each histogram.}
\label{his}
\vskip-3mm
\end{figure}

While the finite-size scaling of $\langle m\rangle$ is governed by the correlation length $\xi$, 
$\langle m_q\rangle$ should instead be controlled by the U(1) length scale $\Lambda$ \cite{oshikawa}, becoming
large for a system of size $L$ only when $L > \Lambda$. 
Fig.~\ref{his} shows magnetization histograms at $h/J=1$ for $Z_4$ and $Z_8$ systems with $L=4$ and $32$. The angular 
distribution $P(\theta)=\int drrP(r,\theta)$ is also shown. The average radius of the distribution is the magnetization 
$\langle m\rangle$, which decreases with increasing $L$. The anisotropy, on the other hand, increases with $L$. This is 
particularly striking for $Z_8$, where the $L=4$ histogram shows essentially no angular dependence, even though $T$ is very 
significantly below $T_{\rm c}$, whereas there are 8 prominent peaks for $L=32$. Thus, in this case the U(1) length scale 
$4 < \Lambda < 32$. For the $Z_4$ system $T$ is much closer to $T_{\rm c}$ but still some anisotropy is seen for $L=4$; 
it becomes much more pronounced for $L=32$.

\begin{figure}
\includegraphics[width=3.75cm,clip]{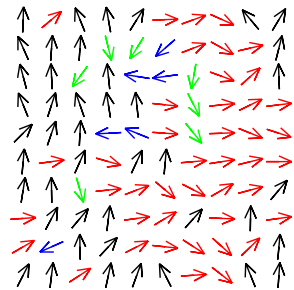}
\caption{(Color online) Spins in one layer of the $Z_4$ model with $L=10$ at $h/J=1,T/J=1.9 < T_{\rm c}$. Here
$m_x \approx m_y$, corresponding to $\theta \approx \pi/4$ in $P(r,\theta)$. Arrows are color-coded according to 
the closest $Z_4$ angle; $n\pi/2$, $n=0,1,2,3$.}
\label{sc}
\vskip-3mm
\end{figure}

It is instructive to examine a spin configuration with $m_x\approx m_y$, i.e., $\theta \approx \pi/4$. Fig.~\ref{sc} shows
one layer of a $Z_4$ system with $L=10$ below $T_{\rm c}$. The spins align predominantly along $\theta=0$ and $\theta=\pi/2$, with 
only a few spins in the other two directions. Clearly there is some clustering of spins pointing in the same direction---the system 
consists of two interpenetrating clusters. Essentially, the configuration corresponds to a size-limited domain wall between 
$\theta=0$ and $\theta=\pi/4$ magnetized states. 

Hove and Sudb{\o} studied the $q$-state clock model and performed a course graining at criticality \cite{hove}. They found that 
the structure in the angular distribution diminished with the size of the block spins for $q \ge 5$, as would be expected if the 
anisotropy is irrelevant. Here we want to quantify the length scale $\Lambda$ at which the anisotropy becomes relevant for 
$T < T_{\rm c}$. Consider first what would happen in a course graining procedure for a single spin configuration of an infinite system 
in the ordered state very close to $T_{\rm c}$. The individual spins will of course exhibit $q$ preferred directions, as is seen 
clearly in Fig.~\ref{sc}, i.e., there would be $q$ peaks in the probability distribution of angles $\theta_i$. Constructing block spins 
of $l^3$ spins, we would expect the angular dependence to first become less pronounced because of the averaging over spins pointing 
in different directions (again, as is seen in Fig.~\ref{sc}). Sufficiently close to $T_{\rm c}$ we would expect the distribution 
to approach flatness. However, since we are in an ordered state, one of the $q$ preferred angles eventually has to become predominant, 
and thus one peak in the histogram will start to grow. This happens at $l \approx \Lambda$. We cannot simulate the infinite system and 
instead we carry out an analogous procedure as a function of the lattice size $L$, sampling a large number of configurations. 
We calculate the order parameters $\langle m\rangle$ and $\langle m_q\rangle$, defined in Eqs.~(\ref{mag},\ref{magstar}), 
and analyze them using
\begin{eqnarray}
\langle m\rangle & = & L^{-\sigma} f(tL^{1/\nu}), \label{mscale} \\
\langle m_q\rangle & = & L^{-\sigma} g(tL^{1/\nu_q}). \label{mqscale}
\end{eqnarray}
Here (\ref{mscale}) is the standard finite-size ansatz with $\sigma=\beta/\nu$, and the XY exponents are $\beta \approx 0.348$ 
and $\nu\approx 0.672$ \cite{compostrini}. Eq.~(\ref{mqscale}) is an intuitive generalization of (\ref{mscale}), which was 
proposed and used also in Ref.~\cite{oshikawa}, but we can actually also derive the scaling function $g(X)$ exactly.

Let us consider the scaling behavior of the order-parameter distribution $P(\vec{m})$. 
It depends upon the system size $L$ and the size of scaling operators that perturb the 
critical theory.  Specifically, we consider the temperature deviation $t = T_{\rm c}-T$ and the 
presumed irrelevant $q$-fold anisotropy strength $h$. By conventional scaling arguments, 
we expect
\begin{equation}
   \label{eq:1}
   P(\vec{m};L,t,h) = L^{\sigma/2} \hat{P}(L^\sigma 
\vec{m},tL^{1/\nu},H=h
   L^{3-\Delta_q}),
\end{equation}
where $\Delta_q>3$ is the scaling dimension of the irrelevant anisotropy.  
The prefactor above is determined from normalization of the
probability distribution.  In the scaling regime, $|t|\ll 1, L\gg 1$, so
$H$ is small.  When the first two arguments are $O(1)$, $\hat{P}$ can be
well-approximated by taking $H=0$ [with ``corrections to scaling'' of
$O(H)$, i.e. suppressed by $L^{3-\Delta_q}$ for a large system].  At
$H=0$, the distribution is fully XY symmetric, and the integral in Eq.~(\ref{magstar})
vanishes.  Thus, in this regime $\langle m_q\rangle$ is small, $O(H)$,
and should be considered as arising from corrections to scaling. This simply 
reflects the irrelevance of the anisotropy at the critical point.

Because the anisotropy is {\sl dangerously} irrelevant, a larger
contribution, however, emerges when $tL^{1/\nu}\gg 1$, i.e.  $L\gg \xi
\sim t^{-\nu}$.  In this limit, the system can be regarded as possessing
long-range XY order, and the only significant fluctuations are the {\sl
global} fluctuations of the XY phase $\theta$.  This is biased by the
anisotropy.  The scale $\kappa$ of the total anisotropy (free) energy
can be estimated by its typical magnitude within an XY correlation
volume, $h\xi^{3-\Delta_q}$, multiplied by the number of correlation
volumes, $(L/\xi)^3$, i.e. $\kappa=h L^3 \xi^{-\Delta_q}$.  Note that
although the energy per correlation volume is small (due to the
irrelevance of anisotropy at the critical point), the number of
correlation volumes becomes very large and more than compensates for
this smallness for $L/\xi$ sufficiently large.

From this argument, we see that for $L/\xi\gg 1$, the distribution of
angles $\theta$ is just determined from a Boltzmann factor for a single
XY spin with the $q$-fold anisotropy energy $\sim -\kappa \cos
q\theta$.  Furthermore, for $L/\xi\gg 1$, the magnitude
$|\vec{m}|\approx \langle m\rangle$ is approximately non-fluctuating.
Thus the distribution factors into the form $P(|m|,\theta) = \langle
m\rangle^{-2} \delta(|m|-\langle m\rangle) P(\theta)$, with
\begin{equation}
   \label{eq:2}
   P(\theta)=
   \frac{1}{Z} e^{ \kappa \cos (q\theta)}.
\end{equation}
Here $Z=\int_0^{2\pi}\! d\theta\, e^{ \kappa \cos (q\theta)}$ is the
single-spin partition function.  It is then straightforward to obtain
from Eq.~(\ref{magstar})
\begin{equation}
   \label{eq:3}
   \langle m_q\rangle = \langle m\rangle \frac{I_1(\kappa)}{I_0(\kappa)},
\end{equation}
where $I_n$ is the modified Bessel function of order $n$. Oshokawa obtained a similar 
expression in a different way, but we disagree with his scaling variable. Comparing
this with the scaling form in Eq.~(\ref{mqscale}), we see that $\nu_q=\Delta_q
\nu/3$ ($a_q=\Delta_q/3$), $\kappa=h (t L^{1/\nu_q})^{3\nu_q}$, and
\begin{equation}
   \label{eq:4}
   g(X) \propto \frac{I_1(\tilde h X^{3\nu_q})}{I_0(\tilde h X^{3\nu_q})}.
\end{equation}
Here $\tilde h$ should be viewed as a non-universal scale factor. From the 
above discussion, one sees that this form is valid for $L/\xi \gg 1$ but $L/\xi_q$ 
arbitrary. For $L/\xi$ of $O(1)$ 
or smaller, $L/\xi_q \ll 1$ (implying $\kappa,X\ll 1$), and the scaling
form for $\langle m_q\rangle$ becomes small and of order the expected correction 
to scaling in the critical regime.

\begin{figure}
\includegraphics[height=10.25cm,clip]{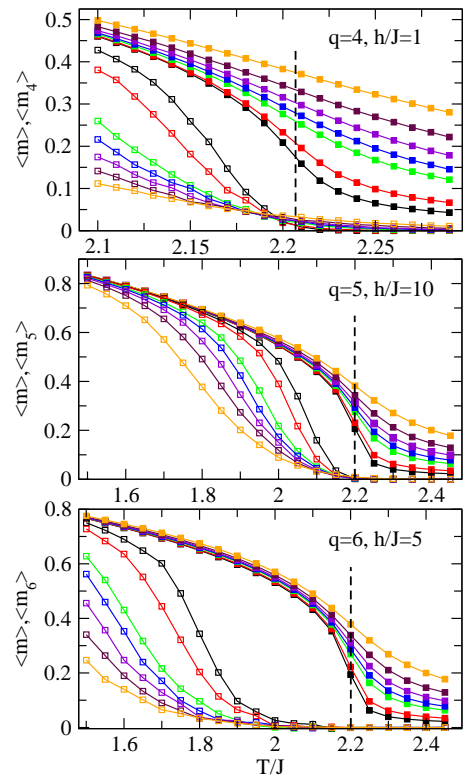}
\caption{(Color online) The XY order parameter $\langle m\rangle$ (solid curves) and the $Z_q$ order parameter $\langle m_q\rangle$ 
(dashed curves) vs temperature for $q=4,5,6$. The system sizes are $L=8,10,12,14,16,24$, and $32$. The curves become
sharper (increasing slope) around $T_{\rm c}$ (indicated by vertical lines). The ratios $h/J$ used are indicated on the graphs.}
\label{mm}
\vskip-3mm
\end{figure}

In Fig.~\ref{mm} we show results for the two order parameters for systems with $q=4,5,6$. We have studied several values of 
$h/J$ and here show results for a different value for each $q$. We have extracted $T_{\rm c}$ using finite-size scaling of $\langle m\rangle$
with Eq.~(\ref{mscale}) and the XY exponents. This works very well for all $q$, confirming the irrelevance of $h$. The magnetization for 
$T < T_{\rm c}$ is seen to decrease marginally with increasing $q$ in Fig.~\ref{mm}. The $Z_q$ order-parameter $\langle m_q\rangle$ changes 
more drastically, being strongly suppressed close to $T_{\rm c}$ for large $q$. This is expected, as $\langle m_q\rangle$ 
should vanish for all $T$ in the XY limit $q\to \infty$. For $Z_4$, the $\langle m_q\rangle$ curves for different $L$ cross 
each other, with the crossing points moving closer to $T_{\rm c}$ as $L$ increases. This is consistent with the above discussion 
of course-graining: In the ordered state close to $T_{\rm c}$, $\langle m_q\rangle$ should first, for small $L$, decrease with 
increasing $L$ as the $q$-peaked structure in $P(\theta)$ diminishes due to averaging over more spins. For larger $L$, $\langle m_q\rangle$ 
starts to grow with $L$ as the length-scale $\Lambda$ is exceeded. This behavior is more difficult to observe directly for $q=5,6$ 
because $\langle m_q\rangle$ is small and dominated by statistical noise close to $T_{\rm c}$ where the curves cross.

\begin{figure}
\includegraphics[height=10.25cm,clip]{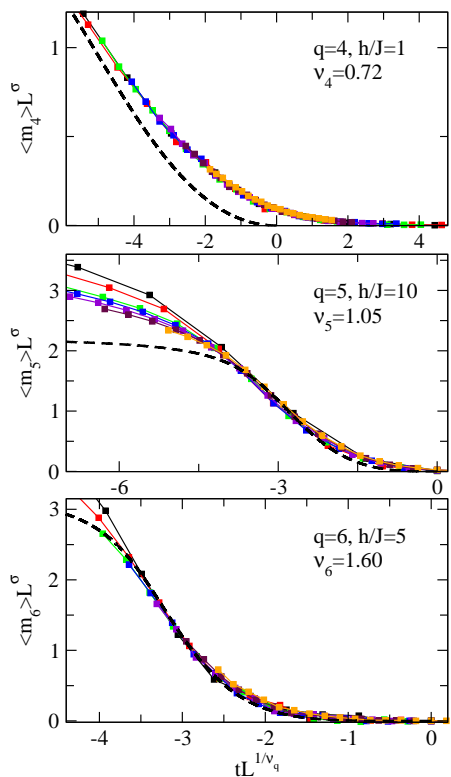}
\caption{(Color online) Scaling of the $Z_q$ magnetization. We use $\sigma=0.52$ for all $q$, 
and $\nu_q$ as indicated in the plots. The colors of the curves correspond to $L$ as in Fig.~\ref{mm}.
The dashed curves are the predicted scaling functions with $\tilde h$ and prefactors adjusted to fit 
the data approximately.}
\label{mszq}
\vskip-3mm
\end{figure}

Fig.~\ref{mszq} shows finite-size scaling of the $Z_q$ order-parameter $\langle m_q\rangle$, using the hypothesis 
(\ref{mqscale}) and the XY value for $\sigma$. Adjusting $\nu_q=a_q\nu$ for each $q$ we find satisfactory data collapse using
$a_4=1.07(3)$, $a_5=1.6(1)$, $a_6=2.4(1)$,  and, not shown in the figure, $a_8=4.2(3)$. These results are consistent with 
the form $a_q = a_4(q/4)^2$, in qualitative agreement with the $\epsilon$-expansion by Oshikawa, which gave $a_q \to q^2/30$ 
for large $q$ \cite{oshikawa}. However, in the $\epsilon$-expansion there are significant deviations from  the $q^2$ form in 
the range of $q$ values considered here. Our $a_6$ is also smaller than the value $\approx 3.5$ obtained on the basis of the 
3-state Potts antiferromagnet \cite{oshikawa}.

In Fig.~\ref{mszq} we also show the scaling function (\ref{eq:4}). It does not match exactly the collapsed 
data, but the agreement improves as $q$ increases. As we have discussed above, the scaling function 
represents the dominant behavior for $T < T_{\rm c}$, but exactly at $T_{\rm c}$ this contribution vanishes and the 
critical-point scaling form $\langle m_q(T_{\rm c}) \rangle \sim L^{3-\Delta_q}$ becomes dominant. For $q=4$, $\Delta_4-3$ is 
small; our estimate is $\Delta_4-3=0.21(9)$, in good agreement with previous estimates of the scaling dimension
\cite{caselle,carmona}. Thus it is clear that very large systems would be required for this contribution to become 
invisible on the scale used in our graph. It is also clear that for $t < 0$, close to $t=0$, there will be a similarly 
significant correction to the asymptotically dominant scaling function. As $q$ increases, we have seen that $\Delta_q=3a_q$ 
increases rapidly, and we thus expect significantly smaller correction to scaling. For $q=6$ the agreement is already 
seen to be quite good, considering that our lattices are not very large. 

To conclude, we relate our results to the quantum VBS states discussed in the introduction. Returning to 
Fig.~\ref{sc}, associating $\theta_i \approx 0$ arrows with two adjacent horizontal dimers on even-numbered columns and 
$\theta_i \approx \pi/2$ with vertical adjacent dimers on even rows, $\langle \theta \rangle=0,\pi/2$ correspond to 
columnar VBS states. A plaquette is a superposition of horizontal and vertical dimer pairs, whence a plaquette VBS 
corresponds to $\langle \theta\rangle=\pi/4$ \cite{levin}. Rotating the arrows by $90^\circ$ corresponds to translating 
or rotating a VBS. Either a columnar or plaquette VBS should obtain in the infinite-size limit, but close to a deconfined 
quantum-critical point, for $L < \Lambda$, the system fluctuates among all mixtures of plaquette and columnar states. 
This corresponds to a ring-shaped VBS order-parameter histogram. In numerical studies of quantum antiferromagnets 
\cite{sandvikvbs,kawashima} no 4-peak structure was observed in the angular distribution, and hence it is not clear what 
type of VBS finally will emerge (although a method using open boundaries favors a columnar state in \cite{sandvikvbs}). 
It seems unlikely that the U(1) symmetry should persist as $L \to \infty$. In the classical $Z_4$ model we never observe a 
perfectly U(1)-symmetric histograms far inside the ordered phase, in contrast to Refs.~\cite{sandvikvbs,kawashima}. On the 
other hand, $a_q$ is larger for $q>4$, and in Fig.~\ref{his} we have shown a prominently U(1)-symmetric histogram for the 
$Z_8$ model deep inside the ordered phase. Thus, the exponent $a$ may be larger for the $Z_4$ quantum VBS than $a_4 \approx 1$ 
obtained here for the classical $Z_4$ model. There is of course no reason to expect them to be the same, as the universality class 
of deconfined quantum-criticality is not that of the classical $Z_4$ model \cite{senthil,sandvikvbs}. Future numerical studies 
of VBS states and deconfined quantum-criticality can hopefully reach sufficiently large lattices to extract the U(1) exponent 
using the scaling method employed here. 

We would like to thank Kevin Beach, Masaki Oshikawa, Andrea Pelissetto, and Ettore Vicari for useful 
discussions and comments. This research is supported by NSF Grant No.~DMR-0513930.

\null

\end{document}